\documentclass[manuscript]{acmart}
\usepackage{algorithm}
\usepackage{algorithmic}
\usepackage{epstopdf}
\usepackage{graphicx}
\usepackage{colortbl}
\usepackage{enumitem}
\usepackage{tikz}
\usepackage{ulem}
\usepackage{multirow}
\usepackage{soul} 
\usepackage{makecell}
\AtBeginDocument{%
  \providecommand\BibTeX{{%
    \normalfont B\kern-0.5em{\scshape i\kern-0.25em b}\kern-0.8em\TeX}}}

\setcopyright{acmcopyright}
\copyrightyear{2018}
\acmDOI{XXXXXXX.XXXXXXX}

\begin{document}

\title{CoHalLo: code hallucination localization based on hidden layer vector mapping}
\title{CoHalLo: code hallucination localization based on hidden layer vector probing}
\title{CoHalLo: code hallucination localization via probing hidden layer vector}

\author{Nan Jia}
\email{jianan\_0101@163.com}
\affiliation{%
  \institution{School of Information Engineering, Hebei GEO University}
  \city{Shijiazhuang}
  \country{China}
  \postcode{050031}
}
\author{Wangchao Sang}
\email{15833588196@163.com}
\affiliation{%
  \institution{School of Information Engineering, Hebei GEO University}
  \city{Shijiazhuang}
  \country{China}
  \postcode{050031}
}

\author{Pengfei Lin}
\email{linpf7@mail2.sysu.edu.cn}
\affiliation{%
  \institution{School of Computer Science and Engineering, Sun Yat-sen University}
  \city{Guangzhou}
  \state{Guangdong}
  \country{China}
  \postcode{519000}
}

\author{Xiangping Chen}
\email{chenxp8@mail.sysu.edu.cn}
\affiliation{%
  \institution{School of Journalism and Communication, Sun Yat-sen University}
  \city{Guangzhou}
  \country{China}
  \postcode{510006}
}

\author{Yuan Huang}
\authornote{Corresponding author.}
\email{huangyuan5@mail.sysu.edu.cn}
\affiliation{%
  \institution{School of Software Engineering, Sun Yat-sen University}
  \city{Zhuhai}
  \state{Guangdong}
  \country{China}
  \postcode{519000}
}

\author{Yi Liu}
\email{liuyi14@pku.edu.cn}
\affiliation{%
  \institution{National Key Laboratory of Data Space Technology and System}
  \city{Beijing}
  \state{Beijing}
  \country{China}
  \postcode{100091}
}

\author{Mingliang Li}
\email{59532499@qq.com}
\affiliation{%
  \institution{School of Information Engineering, Hebei GEO University}
  \city{Shijiazhuang}
  \country{China}
  \postcode{050031}
}

\renewcommand{\shortauthors}{Nan Jia et al.}

\begin{abstract}

The localization of code hallucinations aims to identify specific lines of code containing hallucinations, helping developers to improve the reliability of AI-generated code more efficiently. Although recent studies have adopted several methods to detect code hallucination, most of these approaches remain limited to coarse-grained detection and lack specialized techniques for fine-grained hallucination localization. This study introduces a novel method, called CoHalLo, which achieves line-level code hallucination localization by probing the hidden-layer vectors from hallucination detection models. CoHalLo uncovers the key syntactic information driving the model's hallucination judgments and locates the hallucinating code lines accordingly. Specifically, we first fine-tune the hallucination detection model on manually annotated datasets to ensure that it learns features pertinent to code syntactic information. Subsequently, we designed a probe network that projects high-dimensional latent vectors onto a low-dimensional syntactic subspace, generating vector tuples and reconstructing the predicted abstract syntax tree (P-AST). By comparing P-AST with the original abstract syntax tree (O-AST) extracted from the input AI-generated code, we identify the key syntactic structures associated with hallucinations. This information is then used to pinpoint hallucinated code lines. To evaluate CoHalLo's performance, we manually collected a dataset of code hallucinations. The experimental results show that CoHalLo achieves a Top-1 accuracy of 0.4253, Top-3 accuracy of 0.6149, Top-5 accuracy of 0.7356, Top-10 accuracy of 0.8333, IFA of 5.73, Recall@1\% Effort of 0.052721, and Effort@20\% Recall of 0.155269, which outperforms the baseline methods. 
 
\end{abstract}




\maketitle

\section{Introduction}

In recent years, with the rapid advancement of artificial intelligence technology, large language models (i.e., LLMs) have demonstrated immense potential in code generation \cite{jiang2024survey}. AI tools, such as Copilot \cite{wermelinger2023using} and Cursor \cite{wang2024planning}, have become important programming assistants for developers, automatically generating source code based on demand and significantly boosting development efficiency. However, AI-generated code often exhibits “hallucination” phenomena that do not align with real-world requirements\cite{jang2014survey}. For example, previous studies \cite{spracklen2025we} have shown that the average hallucination rate of the current mainstream LLMs is as high as 20\% to 60\% in code generation. The AI-generated code with hallucinations may be similar to normal code in syntax and style, but hides potential problems, which not only compromises code reliability but also reduces trust in AI tools\cite{ji2024anah}.

Therefore, researchers have focused on the study of the hallucination problem in LLM-based code generation.  Agarwal et al. \cite{agarwal2024codemirage} classified code hallucination into 5 categories, and then they employ LLMs for hallucination detection with zero-shot prompts, while their method exhibits low accuracy in detecting complex hallucinations.  Jiang et al. \cite{jiang2024collu} constructed the Collu-Bench dataset with 13,234 instances, and they employ traditional machine learning and neural network approaches for hallucination detection and the experiments show their best methods (Random Forest and LSTM) only have about 33\% detection accuracy. Tian et al. \cite{tian2024codehalu}  categorized code hallucinations into 4 major types (mapping, naming, resource, logic) with 8 subcategories. They proposed a dynamic hallucination detection method and found that the logical hallucinations are most prevalent.  All these methods did a preliminary exploration of hallucination classification and detection, without further localizing the hallucination.

It is important to locate hallucination from the AI-generated code\cite{rahman2024code}. One common scenario for programmers to generate code using AI tools is first to formulate a natural language requirement, then require the AI tools to generate code based on the requirement, and finally, developers manually determine whether the code meets the actual requirement. In this process, developers need to spend additional time reviewing the code and finding the problems, such as hallucinations from the generated code. If we can automatically determine whether the AI-generated code has a hallucination, and where the hallucination is, it can effectively improve the efficiency of programmers to find and fix the problem\cite {lee2025hallucination}.

There are some challenges in automatically detecting and locating code hallucinations. Firstly, the hallucinating codes may not differ from normal codes, and they can be normally compiled and run, while their implementation logic does not meet the requirements. As a result, it is difficult to identify and locate the hallucinations from the code surface representation.  Secondly, the factors that cause code hallucination are very complex, which may come from the interference of noisy training data, or it may be caused by the inference mechanism of the LLM itself. This concealment and complexity of causing hallucination make it difficult to identify code hallucination from the perspective of its root cause.  

In this study, we propose CoHalLo, which achieves \textbf{co}de \textbf{hal}lucination \textbf{lo}calization by obtaining an effective representation for code hallucination from the hidden layer vector. Specifically, we first fine-tune the pretrained model to identify the hallucinating code. Next, a probe model \cite{hernandez2022ast} is trained to reduce high-dimensional hidden layer vectors to a low-dimensional subspace, which contains the key information to classify the AI-generated code as hallucinated. With the probing technique \cite{hewitt2019structural}, features associated with code hallucinations were amplified, whereas features unrelated to code hallucinations were effectively suppressed. The low-dimensional vectors are used to reconstruct the abstract syntax tree (AST) that represents the code hallucination. Subsequently, we conducted a comparative analysis between the reconstructed AST and the original AST and identified key syntactic elements associated with code hallucinations from the original AST through structural mapping. Finally, the task of locating code hallucinations is achieved by mapping the identified syntactic elements back to their corresponding code lines, assigning confidence scores to each line of code, and sorting them based on these scores.

We conducted experiments on the manually collected code hallucination dataset to evaluate our proposed method. We compare the CoHalLo method with several LLM-based baseline methods, and the experimental results demonstrate that CoHalLo achieves superior performance compared to these baseline methods. Specifically, CoHalLo achieved its best results across different pretrained models, achieving a Top-1 Accuracy of 0.4253, Top-3 Accuracy of 0.6149, Top-5 Accuracy of 0.7356, Top-10 Accuracy of 0.8333, IFA of 5.73, Recall@1\%Effort of 0.052721, and Effort@20\%Recall of 0.155269.

In summary, the main contributions of this paper are as follows:

\begin{itemize}
   \item \textbf{A novel method for locating code hallucinations.} We propose a code hallucination localization technique based on hidden representation detection, which identifies code hallucinations by analyzing the syntactic structures encoded within pretrained models. CoHalLo can extract key features from the hidden layers of the model to achieve fine-grained hallucination localization. To the best of our knowledge, this is the first application of probing techniques in the context of code hallucination localization.
   \item\textbf{The method exhibited good compatibility.} The proposed method does not rely on specific model architectures and is compatible across multiple pretrained code models. We validated its effectiveness across four models: CodeBERT\cite{feng2020codebert}, GraphCodeBERT\cite{guo2020graphcodebert}, UnixCoder\cite{guo2022unixcoder}, and CodeT5\cite{wang2021codet5}, demonstrating its generalization capabilities.
   \item \textbf{Construction of a specialized code hallucination datasets.} We constructed a dataset for code hallucination detection and localization. This dataset not only labels whether the code exhibits hallucinations but also provides a detailed line index for hallucination location. 
   \item \textbf{Comprehensive performance evaluation.} We conducted a systematic evaluation of CoHalLo on our manually collected code hallucination dataset, measuring its performance using multiple evaluation metrics. The experimental results demonstrate that CoHalLo performs well in terms of hallucination detection and localization accuracy.
\end{itemize}

The remainder of this paper is organized as follows:  Section \ref{Approach} provides a detailed introduction to our methodology. Section \ref{Experiment Setup} describes the experimental setup of the study. Section \ref{Experiment Result} presents a detailed analysis of the experimental results of the study. Section \ref{Threats to validity} presents the threats to the validity of this study. Section \ref{Related Work} reviews the related work. Finally, Section \ref{Conclusion and future work} concludes the paper.

\section{Approach} \label{Approach}

\begin{figure*}[htbp]
\centering
\begin{tikzpicture}
\node[fill=white, inner sep=0pt] {\includegraphics[width=1\textwidth]{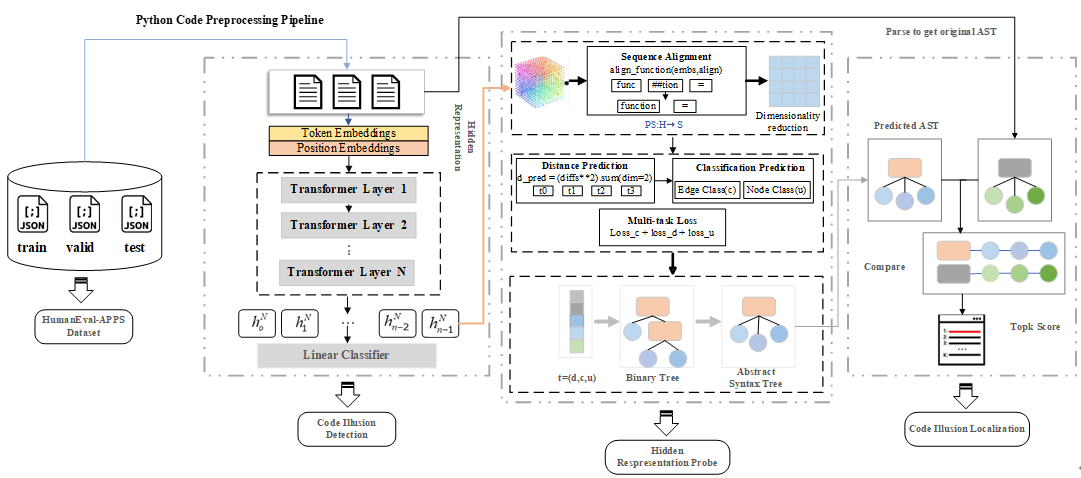}};
\end{tikzpicture}
\caption{\label{fig:figure1}The overall framework of CoHalLo}
\end{figure*}

In this section, we introduce CoHalLo, and Figure \ref{fig:figure1} presents the overall framework of the proposed method. CoHalLo implementation comprises three main stages: 1) code hallucination detection, 2) probing AST mapping and construction, and 3) code hallucination localization. 

\begin{itemize}
   \item \textbf{Code hallucination detection.} As the first stage of hallucination localization, we employ pretrained code generation models (such as CodeBERT) to encode the AI-generated source code and generate a latent representation of semantic information. 
   Subsequently, an additional linear classifier determines whether hallucinations exist in the code. This stage not only provides fundamental hallucination detection capabilities but also offers latent representation information for subsequent hallucination localization analysis.

   \item \textbf{Probing AST mapping and construction.}At this stage, a probe network acts as a specially designed structural decoder. It transforms the high-dimensional latent representations from the pre-trained code model into a low-dimensional subspace specialized for syntax through dimensionality reduction. This process concentrates the syntactic information previously dispersed across neural representations. Within this subspace, the probe network captures complementary syntactic clues by analyzing relational patterns between adjacent token representations: quantifying hierarchical distances between tokens, identifying functional types of syntactic branch nodes to annotate internal nodes, and determining each token's syntactic role to define leaf node properties. Subsequently, the probe network employs a divide-and-conquer strategy to recursively assemble these distributed relational predictions into a hierarchical binary tree structure. By progressively identifying the deepest segmentation points, it partitions the sequence into nested sub-structures, effectively reversing the relational representation of a linear sequence back into a combinatorial tree structure. The generated tree structure is normalized before undergoing structural matching against the original syntax tree. 
   \item \textbf{Code hallucination localization.}In the final stage, we identify key syntactic information related to code hallucinations by comparing the predicted AST (P-AST) and the original AST (O-AST). Compared to the original O-AST, P-AST retains only a subset of the complete syntactic structure, with the hidden representation encoding only limited syntactic information relevant to hallucination localization. By mapping these elements back to their corresponding lines of code, we assigned a confidence score to each line and ranked the lines based on these scores, ultimately achieving code hallucination localization.
\end{itemize}

\subsection{Code Hallucination Detection}

CoHalLo employs a transformer-based\cite{jain2019attention} pretrained model for code hallucination detection. First, we detect code hallucinations by attaching a linear classifier to the selected pretrained model and fine-tuning it on a manually annotated dataset. The model consists of 12 stacked encoder layers pretrained on a large-scale code hallucination dataset to learn general representations of code syntax structures, semantic relationships, and more. The encoder architecture employs a hierarchical-feature extraction strategy. Each encoder layer progressively integrates and refines the features extracted from the input through self-attention mechanisms and feedforward networks, forming a multilevel code representation. Simultaneously, the self-attention mechanism captures long-range dependencies by computing the correlations between each token and all other tokens. For each position in the input sequence, the model can compute the query vector \textit{Q}, key vector \textit{K}, and value vector \textit{V}. The formula for calculating the attention weights\cite{niu2021review} is shown in Equation (1):

\vspace{-0.2cm}
\begin{equation}
\text{Attention}(Q,K,V) = \text{softmax}\left(\frac{QK^T}{\sqrt{d_k}}\right)V
\label{eq:attention}
\end{equation}
\vspace{-0.2cm}

Where $d_k$ is the dimensionality of the key vector, used to scale the pointwise sum of scalars, enabling the model to dynamically focus on other tokens most relevant to the current token, thereby constructing rich contextual representations. To enhance the model's expressive power, multiple independent attention heads were employed in parallel within the encoder layer. This multi-head self-attention mechanism enables the model to learn distinct subspace representations and to capture various aspects of the input. The computation for each attention head is described in Equation (2):

\vspace{-0.2cm}
\begin{equation}
head_i = \text{Attention}(QW_i^Q, KW_i^K, VW_i^V)
\label{eq:multihead}
\end{equation}
\vspace{-0.2cm}

 
The final output of the multi-head is shown as Equation (3):

\vspace{-0.2cm}
\begin{equation}
\text{MultiHead}(Q,K,V) = \text{Concat}(head_1, \ldots, head_h)W^O
\label{eq:multihead_output}
\end{equation}
\vspace{-0.2cm}

This process effectively integrates features from different subspaces into a latent representation, enabling the model to simultaneously focus on various types of code relationships, such as syntactic dependencies, semantic associations, and structural hierarchies.

Simultaneously, within each encoder layer, the position feedforward network (FFN)\cite{soydaner2022attention}  applies a nonlinear transformation to further refine the feature representation, as detailed in Equation (4):

\vspace{-0.2cm}
\begin{equation}
FFN(x) = \max(0, xW_1 + b_1)W_2 + b_2
\label{eq:ffn}
\end{equation}
\vspace{-0.2cm}

The \textit{FFN} enhances the nonlinear expressive power of the model by mapping features to a higher-dimensional representation space through two layers of linear transformations and ReLU activation functions. This refined concealment mechanism enables the model to focus more intently on identifying inconsistencies, logical errors, and other hallucination issues within the code. It encodes the key features required for code hallucination detection and provides inputs for our probe network. By analyzing the syntactic structures and semantic information in these representations, we can identify the underlying factors affecting code hallucination detection, thereby enabling the localization of code hallucinations.

\subsection{Probing AST Mapping and Construction}

At this stage, we denote the latent representation of the code hallucination detection model as $h$, which encodes the fundamental information for hallucination identification and detection\cite{skean2025layer}. To reveal what is encoded in $h$ and how it influences the detection outcomes, we introduce the probe method. This approach facilitates the examination of the features captured by the model, providing deeper insights into the reasoning behind its predictions.

Let $H$ denote the set of all hidden vector representations $h$, where each $h \in H$. The hidden representation contains rich information, but its content is complex owing to the inclusion of irrelevant details. To prevent these extraneous elements from interfering with the results, the probe model specifically focuses on syntactic information within the hidden representation. We assume the existence of a syntactic subspace $S$ encoding AST information. Our objective is to learn a $P_S$ projection that maps the latent representation space $H$ onto the syntactic subspace $S$:

\vspace{-0.2cm}
\begin{equation}
P_s : H \to S
\label{eq:projection}
\end{equation}
\vspace{-0.2cm}

Directly mapping hidden representations to ASTs is challenging because ASTs are not simple vectors but possess a hierarchical structure\cite{latif2023comparison}. To address this, we first transformed ASTs into structured vector representations, thereby establishing a direct learning objective d$P_S$. Lopez et al\cite{hernandez2022ast} verified the existence of syntactic subspaces and established a mapping between ASTs and vector tuples $t = (d, c, u)$, where \textit{d}, \textit{c}, and \textit{u} collectively represent the AST structure.

Following the provisions outlined in the relevant overview, the bidirectional conversion between AST and vector tuples $t=(d,c,u)$ is illustrated in Figure \ref{fig:figure2}. This conversion was performed in two steps:

\begin{figure*}[htbp]
\centering
\includegraphics[width=0.7\textwidth]{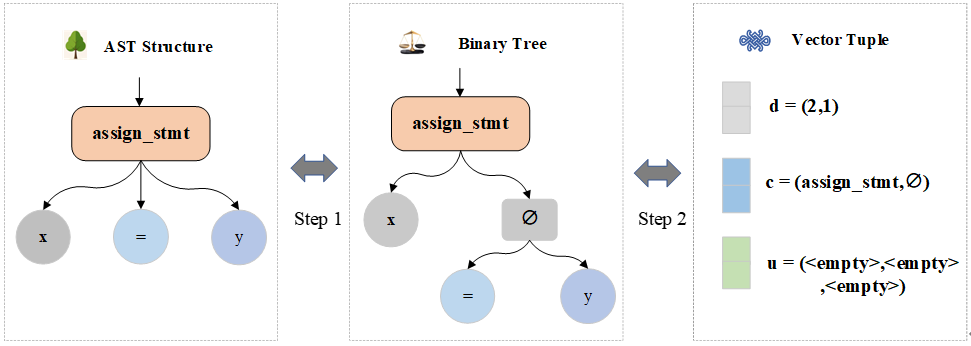}
\caption{\label{fig:figure2}Specific Conversion Process. Rounded rectangles represent non-terminal nodes, and circles represent terminal nodes.}
\end{figure*} 

Step 1: AST to Binary Tree Conversion. To convert an AST into a binary tree, we need to binarize both unary and n-ary nodes. This binary conversion process must satisfy the following three conditions:

1) If a node is n-ary, a special node $\emptyset$ is inserted to binarize it. As shown in Figure \ref{fig:figure2}, the non-terminal node if\_statement in a 3-ary structure is converted to a binary structure by adding the following nodes $\emptyset$.

2) A terminal node (one with only one child node) should merge with its child node to form a new node. This rule applies to both terminal and non-terminal nodes.

3) To convert the binary tree back into an AST, $\emptyset$ must be removed, and the tree structure is reconnected. In addition, previously merged unary nodes must be re-expanded.

Step 2: Binary Tree to Vector Tuple Conversion. Next, we convert the binary tree into a tuple $t=(d,c,u)$, completing the two-step transformation from the AST to the vector tuple representation.

1) \textit{d} encodes the structural information. The \textit{d} vector records the structural relationships between adjacent terminal nodes in a binary tree. Each $d_i$ value represents the path length from the root node to the lowest common ancestor of two adjacent terminals $(w_i,w_i+1)$, reflecting the hierarchical relationship between the two terminals within the syntax tree. For example, $d=(2,1)$ indicates that the common ancestor of the first and second terminals is at the second level, whereas the common ancestor of the second and third terminals is at the first level.

2) \textit{c} encoding tag information. The \textit{c} vector stores the syntactic labels of the lowest common ancestor for each pair of adjacent terminals, reflecting the syntactic structure type of the code. The dimension of \textit{c} is identical to that of the \textit{d} vector, as each pair of adjacent terminals corresponds to a single common ancestor. For example, $c=(if\_statement,\emptyset)$ indicates that the common ancestor of the first pair of terminals is the if\_statement node, whereas the second pair is the empty node.

3) \textit{u} encodes label information. The \textit{u} vector records whether each terminal node merges with a non-terminal node. If a terminal node merges with a non-terminal node, \textit{u} records the merged label; if no merger occurs, it records <empty>. The dimension is the number of terminals \textit{n}.

Using $t=(d,c,u)$, the corresponding AST can also be obtained through reverse conversion.

\subsection{Code Hallucination Localization}

P-AST is an abstract syntax tree extracted and reconstructed from latent representations, embodying the encoded syntactic information that directly influences the decision-making process of code hallucination detection models. During fine-tuning, features relevant to code hallucinations are enhanced, whereas those irrelevant to code hallucinations are suppressed. The Original Abstract Syntax Tree (O-AST) is a genuine abstract syntax tree constructed directly from the source code using parsers such as Tree-Sitter, representing the actual syntactic structure of the code. Based on this understanding, we hypothesize that changes in syntactic features embedded within the latent representation will lead to corresponding alterations in the extracted P-AST, which are closely tied to the decision-making of code hallucination models. In this section, we achieve code hallucination localization by analyzing the differences between the P-AST and O-AST. We compared the two AST approaches and provided pseudocode for the AST comparison and scoring processes. By comparing the discrepancies between P-AST and O-AST, the system can identify deviations in the syntactic understanding of the model, thereby detecting code hallucinations and enabling precise line-level localization.

\subsubsection{Comparison.} 

To analyze the syntactic information retained in the latent representation, we extracted P-AST from the model's latent states and compared it with O-AST parsed directly from the source code. This comparison revealed the key syntactic structures that the model deemed important for code hallucination detection. To compare P-AST and O-AST, suitable representations for AST substructures must first be established. López et al.\cite{hernandez2022ast} employed tree component representations to assess the degree of overlap between AST. The tree component representations are defined as:

\vspace{-0.2cm}
\begin{equation}
\langle \text{syntax tree root node} \rangle - (i) - (i+1) - \ldots - (j)
\label{eq:tree_component}
\end{equation}
\vspace{-0.2cm}

Here, $i, i+1, \ldots, j$ denotes the token ID of the input code. This representation captures the root node and all the leaf nodes of the subtree while omitting its internal structure. To obtain a more complete representation of each subtree, we perform a preorder traversal and derive the representation as follows:

\vspace{-0.2cm}
\begin{equation}
\langle \text{subtree root node} \rangle - \langle \text{preorder traversal} \rangle - \ldots
\label{eq:subtree_representation}
\end{equation}
\vspace{-0.2cm}

Unlike tree component representations that record only roots and leaves, preorder traversal representations\cite{valiente2002algorithms} preserve the hierarchical structure of subtrees more comprehensively. This structured representation better captures the key syntactic patterns used for comparison.

\subsubsection{Scoring.}  After comparing the AST, we identified the syntactic structures preserved in the latent representations. We consider the lines of code containing these structures to be highly relevant to hallucinations. To facilitate hallucination localization, we assigned scores to the tokens corresponding to matching subtrees, reflecting their importance in the model's prediction process for hallucination detection. After completing the scoring process, we summed the token scores for each line of code to calculate the line score. The lines of code are then sorted by score, with higher-scoring lines deemed more relevant to the hallucination, thereby completing the hallucination localization.

The scoring process primarily consists of two stages: the first stage involves token-level statistical scoring\cite{ren2020codebleu}. The system traverses the predicted multiple structures (pred\_multiset) and compares each structural node with the actual label. If the prediction was correct, a base score of 1 was awarded. For control flow-related syntactic structures (such as if statements, loop statements, and exception handling), a weighted score of 1.5 points was assigned. The second stage involved line-level score aggregation, mapping token-level scores to code line-level scores using correlation functions. For strictly aligned cases, the line-level score was directly calculated by summing the scores of tokens corresponding to each line. For misaligned cases, the system attempts to re-perform token segmentation and mapping on the input data. This ultimately generates a lines\_score, which contains the score for each line of code. Lines with higher scores are considered more relevant to the code hallucination, thereby completing the localization of code hallucinations from the token level to the line level.

The scoring process is shown in Algorithm 1. First, lines 1-3 perform the initialization preparation phase: Line 1 extracts all syntactic structures from the predicted AST (P-AST) and stores them in structuresP. Line 2 extracts all syntactic structures from the original AST (O-AST) and stores them in structuresO. Line 3 initializes the SCORE\_VECTOR array with a zero vector of length equal to the number of tokens in the input. Starting from Line 4, iterate through the syntax structures in the predicted AST, examining each structure in the P-AST one by one. Line 5 performs structure matching verification by checking whether the current predicted structure exists in the original AST structure set `structuresO'. Line 6 retrieves the root node type of the matched structure to prepare for the weight allocation. Lines 7-11 begin processing the weight allocation related to the control flow. Line 7 determines whether the root node type is a control flow structure. Line 8 assigns a higher control weight (CONTROL\_WEIGHT, 1.5 points) to the control flow structures. Line 10 assigns a base weight (BASE\_WEIGHT, 1.0 points) to the non-control flow structures. Line 12 extracts all token indices associated with the current structure to determine the specific token positions that require scoring. Lines 13-15 perform weighted scoring on relevant tokens. Line 13 iterates through each associated token index in the current structure. Line 14 accumulates the corresponding weight value at the appropriate position in the SCORE\_VECTOR. Accumulation is used because a single token can simultaneously belong to multiple syntactic structures. If no matching structure exists in the O-AST, the structure is skipped without scoring.

\begin{algorithm}
\caption{Token-Level Scoring with Control Flow Weighting}
\label{alg:token_scoring}

\begin{flushleft}
\textbf{Input:} P-AST (Predicted AST), O-AST (Original AST) \\
\textbf{Output:} TOKEN\_SCORE
\end{flushleft}

\begin{enumerate}[label=\arabic*:]
\item extractStructures(P-AST) $\rightarrow$ structuresP
\item extractStructures(O-AST) $\rightarrow$ structuresO
\item initialize TOKEN\_SCORE $\leftarrow$ [0, 0, \ldots, 0]
\item \textbf{for each} structure s $\in$ structuresP \textbf{do}
\item \hspace{1em} \textbf{if} s $\in$ subtreesO \textbf{then}
\item \hspace{2em} nodeType $\leftarrow$ getRootType(s)
\item \hspace{2em} \textbf{if} nodeType $\in$ CONTROL\_WEIGHT \textbf{then}
\item \hspace{3em} weight $\leftarrow$ CONTROL\_WEIGHT $\cdot$ 1.5
\item \hspace{2em} \textbf{else}
\item \hspace{3em} weight $\leftarrow$ BASE\_WEIGHT $\cdot$ 1.0
\item \hspace{2em} \textbf{end if}
\item \hspace{2em} tokenIndices $\leftarrow$ getTokenIndices(s)
\item \hspace{2em} \textbf{for each} idx $\in$ tokenIndices \textbf{do}
\item \hspace{3em} TOKEN\_SCORE[idx] $\leftarrow$ TOKEN\_SCORE[idx] + weight
\item \hspace{2em} \textbf{end for}
\item \hspace{1em} \textbf{end if}
\item \textbf{end for}
\item \textbf{return} TOKEN\_SCORE
\end{enumerate}

\end{algorithm}

Finally, we aggregated the scores obtained from the annotations to calculate the score for each line of code. Specifically, for each line of code, we summed the scores of all annotations within that line to compute the cumulative score.

\section{Experiment Setup} \label{Experiment Setup}

In this section, we introduce the experimental setup. We first introduce the evaluation metrics adopted. Next, we detail the datasets used in the experiments. 
All experiments were conducted on a vGPU-32GB server equipped with a 12vCPU Intel® Xeon® Platinum 8352V CPU @ 2.10GHz and 32 GB of physical memory.

\subsection{Evaluation Metrics}

\textbf{F1 Score (F1):} We employed the F1 score, precision, and recall as the evaluation metrics to measure each model's ability to identify code hallucinations. The F1 score balances precision and recall, providing a comprehensive reflection of the classification performance of the model. 
Its calculation formula is given by Equation:

\vspace{-0.2cm}
\begin{equation}
F1 = 2 \times \frac{Precision \times Recall}{Precision + Recall}
\label{eq:f1_score}
\end{equation}
\vspace{-0.2cm}

Precision and Recall are expressed as shown in the following Equations:

\vspace{-0.2cm}
\begin{equation}
Precision = \frac{TP}{TP + FP}
\label{eq:precision}
\end{equation}
\vspace{-0.2cm}

\vspace{-0.2cm}
\begin{equation}
Recall = \frac{TP}{TP + FN}
\label{eq:recall}
\end{equation}
\vspace{-0.2cm}

Precision and recall are calculated by comparing the model's predicted hallucination labels with the true labels. \textit{TP} denotes correctly identified hallucination samples, \textit{FP} denotes normal samples misclassified as hallucinations, and \textit{FN} denotes missed hallucination samples.

\textbf{Top-k Accuracy (Top-1/3/5/10 ACC):} Top-k accuracy serves as a key metric for evaluating a model's precision in locating the code hallucinations. Specifically, Top-1 indicates whether the model's top-ranked prediction contains a code hallucination, and Top-3 signifies that at least one of the top three predictions includes a code hallucination. Top-5 indicates whether at least one of the top five predicted positions contains code hallucinations, and Top-10 indicates whether at least one of the top ten predicted positions contains code hallucinations. This metric reflects the model's ability to identify code hallucinations within a limited inspection scope. Higher values indicate the model's greater capability to accurately locate code hallucinations at fewer positions.

\textbf{Initial False Alert (IFA):} IFA refers to the first instance in which the model incorrectly identifies a code segment as containing code hallucinations during the initial detection, despite the segment being legitimate. This metric measures the accuracy of the model's “first impression” during detection. A lower IFA value indicates that the model is less prone to misclassification early in the detection process. This is crucial for user experience and development efficiency in practical applications, as it reflects the model's ability to initiate code hallucination detection quickly and accurately.

\textbf{Recall@1\%Effort (R@1\%E):} R@1\%E refers to the proportion of code hallucinations that the model can successfully identify while inspecting only 1\% of the total code volume. This metric measures the detection efficiency of the model under extremely low inspection costs. A higher value indicates that the model can uncover more code hallucinations with less effort.

\textbf{Effort@20\% Recall (E@20\%R):} E@20\%R refers to the percentage of total code that a model must examine to achieve a 20\% recall rate for code hallucinations. This metric measures the computational effort required for the model to achieve a specific detection target. A lower value indicates that the model can achieve the desired detection performance with less computational effort than other models.

\subsection{Dataset}

Although existing studies \cite{agarwal2024codemirage, jiang2024collu} have collected code hallucination datasets, none of these datasets are annotated with the specific line of code where the hallucination occurs, and they only provide the sample-level hallucination label. Therefore, these datasets are not suitable for the hallucination localization task in this work. Then, we constructed a self-built dataset based on the two standard datasets of HumanEval\cite{li2024humaneval} and APPS\cite{hendrycks2021measuring}.
During dataset construction, we employ 6 language models (i.e., CodeLlama-7B, DeepSeek-Coder-16B, Qwen2.5-Coder-32B, DeepSeek-V2.5, GPT-4o-2024, and Claude-3.5-Sonnet) to generate the answers for 500 programming problems from APPS and HumanEval. For each programming problem, each large model generated 5 answers. Consequently, we collected 15,000 answers (each answer corresponding to a code sample). We invited 20 annotators (with an average of over three years of programming experience) to validate each answer. Every two annotators form a group for independent annotation. If annotation disagreement occurs, the third annotator verifies the result. Based on the manual verification results, 6,683 answers (i.e., code samples) were annotated as hallucinated at the line of code level. 

To construct the hallucination taxonomy, we first performed open coding on a sample of these hallucinated answers. During this stage, annotators independently labeled plausible error types and provided detailed notes on the error’s cause. Subsequently, all annotators convened to review, discuss, and consolidate the collected labels. Overlapping categories were merged, and any discrepancies were resolved through rigorous, in-depth discussion to iteratively construct and refine the final taxonomy. Finally, we summarized 7 types of coarse-grained hallucinations and 13 types of fine-grained hallucinations.




\section{Experiment Result} \label{Experiment Result}

In this section, we pose the following four research questions: 

RQ1: How do different models perform in hallucination classification? 

RQ2: How do different pretrained models affect CoHalLo's localization performance? 

\subsection{RQ1: How do different models perform in hallucination classification?}

\subsubsection{\textbf{Motivation.}} In the first step of our proposed method, we need to determine whether an AI-generated code has hallucination, i.e., conducting a binary classification task. Then, 
For the hallucinating sample identified by the model, its hidden layer vector is obtained from the classification model and then used for the next step of hallucinating localization.
Considering that different classification models may exhibit significant performance variations in this binary classification task due to differences in their architectural design, training strategies, and optimization objectives,  we selected four pretrained models (UnixCoder, CodeBERT, GraphCodeBERT, and CodeT5) for comparison. Through the hallucination detection accuracy, we aim to identify the model that best suits the binary classification task, and further locate the code lines where the hallucination occurs via probing the hidden layer vectors in the next step. 



\subsubsection{\textbf{Design.}}

For all the pretrained model,  we employed identical data preprocessing workflows and  training strategies to ensure a fair comparison of different model for the hallucination detection task. 
We split the dataset into training, validation, and testing sets at an 8:1:1 ratio for subsequent experiments. During the hallucination detection phase, we first perform full fine-tuning on the pre-trained models to better adapt them to the task. Data processing involves tokenizing the filtered data using RobertaTokenizer, truncating each data point to 510 tokens, inserting [CLS] and [SEP] special tokens, and finally padding the sequence length uniformly to 512 tokens. Simultaneously, we built a classification head on top of the pre-trained encoder. This head consists of a fully connected layer, Dropout, and a binary classification output layer for hallucination classification. Training employed the AdamW optimizer, with each model running independently for 20 epochs. Cross-entropy loss and gradient clipping were used to ensure training stability.

\subsubsection{\textbf{Results.}} Table~\ref{tab:classification_performance} presents the performance of the competing models in the code hallucination classification task. CodeBERT achieves the best precision, and GraphCodeBERT achieves the best recall and  F1 score. Meanwhile, we found that the F1 scores achieved by the four pretrained models have no significant difference, and all four pretrained models achieved F1 scores exceeding 0.83. The results demonstrate that models pretrained on large-scale code corpora can understand code semantics, thereby achieving similar performance in the hallucination classification task.

\begin{table}[h]
\centering
\caption{Hallucination Classification Performance of Different Models}
\label{tab:classification_performance}
\begin{tabular}{lcccc}
\toprule
\textbf{Models} & \textbf{Precision} & \textbf{Recall} & \textbf{F1} \\
\midrule
CodeBERT & \textbf{0.8889} & 0.8144 & 0.850  \\
UnixCoder & 0.8466 & 0.8263 & 0.8364 \\
GraphCodeBERT & 0.8623 &\textbf{ 0.8623} &\textbf{ 0.8623} \\
CodeT5 & 0.8734 & 0.8263 & 0.8492 \\
\bottomrule
\end{tabular}
\end{table}


\subsection{RQ2: How do different pretrained models affect CoHalLo's localization performance?}

\subsubsection{\textbf{Motivation.}} The probe method is model-agnostic, meaning that it can be applied to any pretrained model capable of generating latent representations for code. In this research question, we systematically replaced different pretrained models to investigate the impact of model selection on the overall performance of CoHalLo in the task of code hallucination localization. We specifically analyzed the multiple metrics, such as Top-K accuracy, IFA, to evaluate the performance of CoHalLo. 

\subsubsection{\textbf{Design.}} To answer this research question, we employed the four pretrained models (UnixCoder, CodeBERT, GraphCodeBERT, and CodeT5) as foundational components of the CoHalLo framework. These models are tasked with performing identical code hallucination detection, from which we extract the corresponding latent vectors to serve as inputs for the probe model, and CoHalLo performs code hallucination localization based on the probing results. In this RQ, we use the evaluation metrics of Top-k accuracy ($k{=}1,3,5,10$), IFA (Inspection for Accuracy), R@1\%E (Recall at 1\% Effort), and E@20\%R (Effort at 20\% Recall) for result analysis. 




\subsubsection{\textbf{Results.}} Table \ref{tab:localization_analysis} presents the performance metrics of CoHalLo when using different code hallucination detection models. Overall, all models exhibit a consistent tendency as the Top-k value increases: localization accuracy steadily rises, indicating that hallucinated code lines are ranked higher in the output. These results demonstrate that all four pretrained models achieve satisfactory performance in code hallucination localization. 


\begin{table}[h]
\centering
\caption{Code Hallucination Localization Results Across Different Models}
\label{tab:localization_analysis}
\begin{tabular}{lccccccc}
\toprule
\textbf{Models} & \textbf{Top-1} & \textbf{Top-3} & \textbf{Top-5} & \textbf{Top-10} & \textbf{IFA} & \textbf{R@1\%E} & \textbf{E@20\%R} \\
\midrule
CodeBERT & 0.4253 & 0.5747 & 0.6897 & 0.7644 & 6.47 & \textbf{0.057823} & 0.156865 \\
UnixCoder & 0.4138 & 0.5977 & 0.7299 & \textbf{0.8448} & 5.77 & 0.052721 & 0.163245 \\
GraphCodeBERT & \textbf{0.4253} &\textbf{ 0.6149 }& \textbf{0.7356} & 0.8333 &\textbf{ 5.73} & 0.052721 & 0.155269 \\
CodeT5 & 0.3793 & 0.5632 & 0.6782 & 0.8103 & 5.86 & 0.05102 & \textbf{0.152209} \\
\bottomrule
\end{tabular}
\end{table}



We also observed that different models display slight variations in code hallucination localization. GraphCodeBERT achieves the best results in terms of Top-1$\sim$Top-5 accuracy and IFA in code hallucination localization. This may be attributed to its graph structure modeling mechanism that enables GraphCodeBERT to gain a deeper understanding of the semantic relationships and contextual dependencies of the code\cite{guo2020graphcodebert}. Meanwhile, we found that the accuracy achieved by some models has no significant differences, such as CodeBERT and UniXcoder. Given the better results exhibited by the GraphCodeBERT, we will use it as the base model for CoHalLo in the subsequent RQs.


\section{Threats to validity} \label{Threats to validity}




\textbf{Internal Validity Threats.}
To ensure the reliability of the experimental results, we adopted a unified training workflow and standardized evaluation protocols. However, random initialization during probe model training, data sampling strategies, and the stochastic nature of the optimization algorithms may introduce variability. Although different strategies utilize the same base model, variations may arise in the initialization and training processes of probe layers. To minimize such discrepancies, we employed default hyperparameters for each method, ensuring the correctness and consistency of our settings.


\textbf{External Validity Threats.}
Our experiments were validated on a specific, manually annotated code dataset in which the probe method demonstrated structural comprehension capabilities. However, the diversity of programming styles and code organization across different software projects may affect the generalization performance of the method. Current evaluations primarily focus on specific Python code patterns, and further validation is needed regarding their adaptability to emerging programming languages, specialized programming paradigms, and large-scale software systems. Future work will expand the scope of experiments to include more diverse programming languages, code repositories across different domains, and code snippets from real-world development scenarios, thereby further enhancing the quality and breadth of datasets.

\section{Related Work} \label{Related Work}






With the remarkable success of LLM-based methods in code generation tasks, researchers have increasingly focused on the reliability of content produced by LLMs. LLMs may generate syntactically correct but semantically erroneous code snippets (i.e., a phenomenon known as ``code hallucination''), which severely undermines the models' credibility and practicality in real-world development. Consequently, research addressing hallucination issues in large-scale model-generated code has begun to emerge. 

Agarwal et al.~\cite{agarwal2024codemirage} categorized code hallucinations into five types and employed zero-shot prompting to detect hallucinations using large models. However, their approach only classified hallucinations during detection and encountered issues when handling complex hallucinations. 
Jiang et al.~\cite{jiang2024collu} constructed the Collu-Bench dataset with 13,234 instances, and employed traditional machine learning and neural network approaches for hallucination detection. Experiments demonstrated that LSTM (33.15\%) achieved the best performance. 
Tian et al.~\cite{tian2024codehalu} categorized code hallucinations into four main types (mapping, naming, resource, and logic) and eight subcategories, proposing a dynamic hallucination detection method and finding that logical hallucinations were the most prevalent. 
 Zhang et al.\cite{zhang2025llm} were the first to study hallucination in repository-level code-generation tasks; employing several models on the CoderEval\cite{yu2024codereval} dataset, they carried out generation experiments, formulated a hallucination taxonomy for repository-level code, and analyzed possible causes and mitigation strategies.

Although these methods represent preliminary explorations into hallucination detection, they remain confined to identifying whether code samples contain hallucinations without further pinpointing their precise locations. Addressing this critical issue of code hallucination localization, we proposed the CoHalLo method, which supports multiple pretrained models, including CodeBERT, UnixCoder, and CodeT5, analyzing their hidden representations through probes. This structural probing method enables us to identify syntactic comprehension deficiencies in code generation, offering a viable solutions for code hallucination localization.

\section{Conclusion and future work} \label{Conclusion and future work}


In this study, we propose CoHalLo, which identifies and localizes code hallucinations by probing the model's hidden representations. CoHalLo reconstructs the AST structure from these hidden representations and compares it against the actual AST derived from the source code. By analyzing discrepancies in structural reconstruction, CoHalLo identifies the code where the model exhibits understanding bias, enabling precise hallucination localization. 
Experimental results demonstrate that the probe-based detection method exhibits significant advantages in code hallucination localization tasks. 
In our future work, we plan to expand the capabilities of code hallucination detection and localization. For example, we will explore extracting richer code attributes, such as semantic information and data flow dependencies from latent representations to build a more comprehensive code comprehension framework.  Furthermore, we will investigate the generalization capabilities of hallucination detection and localization across programming languages.




\bibliography{software}










\end{document}